\documentclass[12pt,a4paper]{article}
\usepackage{amsmath}
\usepackage[dvips]{graphicx}
\input epsf
\usepackage{times}
\begin{document}
\begin{titlepage}
\title{Effects of reflective scattering in nuclear collisions}
\author{S.M. Troshin, N.E. Tyurin\\[1ex]
\small  \it Institute for High Energy Physics,\\
\small  \it Protvino, Moscow Region, 142281, Russia}
\normalsize
\date{}
\maketitle

\begin{abstract}
 We discuss
effects of reflective scattering for heavy nuclei collisions and
the deconfined matter formation at the LHC and asymptotical
energies. Reflective scattering in hadron collisions leads to
decreasing matter density with energy beyond the LHC energies.
Limiting form of energy dependence of hadron density is obtained.
\end{abstract}
\end{titlepage}
\setcounter{page}{2}

\section*{Introduction}
Deconfined state of matter has been discovered  by the four major
experiments at RHIC \cite{rhic}. The highest values of energy and
density have been reached in these experiments. This deconfined
state appears to be strongly interacting collective state
with properties of the perfect liquid.   The nature of the
deconfined form of matter is not known.  The importance of the
experimental discoveries at RHIC is that the matter is strongly
correlated and reveals high degree of the coherence when it is
 well beyond the critical values of density and
temperature.

In this note  we address   one aspect of the broad  problem of
transition to the deconfined state of matter, namely, we  discuss
the role of the reflective scattering on the energy dependence of
density in the percolation mechanism of the transition to the
deconfined state of matter.

The main idea of the percolation mechanism of deconfinement is a
formation in the certain volume of a connected hadron cluster due
to increasing temperature and/or hadron density \cite{satz}, i.e.
when vacuum as a connected medium disappears,  the deconfinement
takes place. This process has typical critical dependence on
particle density. Thus, it was proposed to use percolation to
define the states of matter and consider the disappearance of a
large-scale vacuum as the end of hadronic matter
existence\cite{satz,satz2}.

The reflective scattering in its turn is a natural interpretation
of the unitarity saturation based on the optical concepts in high
energy hadron scattering \cite{inta}. At very high energies the
elastic scattering can go beyond the black disk limit and become a
reflective  at small impact parameters, i.e. elastic $S$-matrix
gets negative values.   The evolution with energy  is
characterized by increasing albedo due to the  interrelated
increase of reflection
  and decrease of absorption at small impact parameters.
Asymptotically, picture of  particle collisions with small impact
parameters resembles  hard balls collisions. Therefore, the presence of
reflective scattering reduces the available for scattering volume and results in lower
densities of hadronic matter needed for triggering percolation mechanism
of deconfinement.

\section{Effective degrees of freedom and overlap of hadrons as a mechanism of deconfinement}

For the beginning, we would like to note that there is an interesting possibility that
the origin of the transient state in hadron and nuclear collisions
and its dynamics along with hadron structure can be related to
the mechanism of spontaneous chiral symmetry breaking ($\chi$SB) in QCD \cite{bjorken},
 which  leads
to the generation of quark masses and appearance of quark condensates. This mechanism describes
transition of the current into  constituent quarks.
The  gluon field is considered to be responsible for providing quarks with
  masses and its internal structure through the instanton
  mechanism of the spontaneous chiral symmetry breaking.
Massive  constituent quarks appear  as quasiparticles, i.e. current quarks and
the surrounding  clouds  of quark--antiquark pairs.   Quark  radii are
determined by the radii of  the  surrounding clouds.

  Collective excitations of the condensate are the Goldstone bosons,
and the constituent quarks interact with each other via exchange
of the Goldstone bosons; this interaction is mainly due to pion field.
Pions themselves are the bound states of massive
quarks. For quark-pion interaction
  the following form can be used \cite{diak}:
 \begin{equation}
{\cal{L}}_I=\bar Q[i\partial\hspace{-2.5mm}/-M\exp(i\gamma_5\pi^A\lambda^A/F_\pi)]Q,\quad \pi^A=\pi,K,\eta.
\end{equation}
This interaction is strong. The general form of the total
effective Lagrangian (${\cal{L}}_{QCD}\rightarrow
{\cal{L}}_{eff}$)
 relevant for
description of the non--perturbative phase of QCD
 includes the three terms \cite{gold} \[
{\cal{L}}_{eff}={\cal{L}}_\chi +{\cal{L}}_I+{\cal{L}}_C.\label{ef} \]
Here ${\cal{L}}_\chi $ is  responsible for the spontaneous
chiral symmetry breaking and turns on first.
To account for the
constituent quark interaction and confinement the terms ${\cal{L}}_I$
and ${\cal{L}}_C$ are introduced.  The  ${\cal{L}}_I$ and
${\cal{L}}_C$ do not affect the internal structure of the constituent
quarks.

The picture of a hadron consisting of constituent quarks embedded
 into quark condensate implies that overlapping and interaction of
peripheral clouds   occur at the first stage of hadron interaction.
At this stage,  part of the effective lagrangian ${\cal{L}}_C$ is turned off
and transition to the deconfined state of hadron matter occurs.

The mechanism described above \cite{inte}, when the part of the
effective lagrangian ${\cal{L}}_C$ is suddenly turning off, can be
formulated in terms of percolation theory. As it was already
mentioned in the Introduction, it was recently proposed to use
this theory as a mechanism of deconfinement in the form of a rapid
``cross-over'' in  \cite{satz,satz2}. Usually, phase
transitions are associated with symmetry breaking and
singularities in the partition function. Percolation theories also
describe critical behavior, but those consider geometrical
quantities such as cluster formation (cf. \cite{satz} and
references therein). These quantities also diverge but its
divergence cannot be obtained from the partition function.

\section{Reflective scattering and percolation}
 In \cite{satz} the two possibilities for the deconfinement were considered: one is
related to the percolation of permeable spheres (mesons) and
another one deals with percolation of spheres with repulsive cores
(baryons). While the latter is a good approximation for the static
case and the case of increasing baryon density, it seems to be not
appropriate for the real processes of nuclei collisions. Indeed, the
presence of the repulsive core in the structure of baryons is
essential for preventing nuclei from collapsing. But, in the
processes of nuclear collisions  they loose their identity.
Therefore, no reasons exist to extend static properties to the
dynamics of nuclear interactions and expect presence of the
repulsive core in baryons participating in nuclear collisions.
Indeed, experimental studies of the hadron elastic and inelastic
overlap functions
\[
h_{el}(s,b)\equiv\frac{d\sigma_{el}}{db^2}= {4\pi}|f(s,b)|^2 ;\,\,
h_{inel}(s,b)\equiv\frac{d\sigma_{inel}}{db^2}={4\pi}\eta(s,b)
\]
at {\it{modern energies}} demonstrate an absence of the region at small impact parameters
where probability
of the elastic and/or inelastic overlaps of hadrons (mesons and baryons) is small or zero.
Note here, that the unitarity relation
written for the elastic scattering amplitude $f(s,b)$ in the impact parameter representation relates it
with the contribution of the inelastic channels $\eta(s,b)$:
\begin{equation}
\mbox{Im} f(s,b)=|f(s,b)|^2+\eta(s,b). \label{ub}
\end{equation}

The situation is expected to be different at higher energies when inelastic overlap function
 $h_{inel}(s,b)$ would have a peripheral $b$-dependence and will
tend to zero for $b=0$ at $s\to\infty$ {cf. e.g \cite{inta}).
As it was pointed out in the Introduction, corresponding behavior of elastic scattering
$S$-matrix $S(s,b)$ can then be interpreted
as an appearance of a reflecting ability of scatterer due to increase of
 its density beyond some critical value, corresponding to refraction index  noticeably
 greater than unity. In another words, the scatterer has now not only
 absorption ability (due to  presence of inelastic channels), but it starts to be reflective at very
 high energies. The appearance of this reflective ability in its turn
   can be related to the
 finite size of an extended parton \cite{gpd}. In central collisions, $b=0$, scattering
  approaches to the completely
 reflecting limit $S=-1$ at $s\to\infty$.

 Transition to the reflective scattering mode is naturally
 reproduced by the $U$-matrix form of elastic unitarization.
The elastic scattering $S$-matrix ($2\to 2$ scattering matrix element)
in the impact parameter representation is
written in this unitarization scheme in the form of linear fractional transform:
\begin{equation}
S(s,b)=\frac{1+iU(s,b)}{1-iU(s,b)}, \label{um}
\end{equation}
where $U(s,b)$ is the generalized reaction matrix, which is
considered to be an input dynamical quantity similar to an input
Born amplitude.
 For simplicity consider
 the case of pure imaginary $U$-matrix and make the replacement $U\to iU$, i.e.
\begin{equation}
S(s,b)=\frac{1-U(s,b)}{1+U(s,b)}. \label{umi}
\end{equation}
It can  easily be seen that reflective scattering mode (when $S(s,b)<0$) starts to appear at
the energy $s_R$, which is determined as a solution of the equation
\[
U(s_R,b=0)=1.
\]
At $s>s_R$ the elastic scattering  acquires ability for
reflection, while  inelastic overlap function $h_{inel}(s,b)$ gets a
 peripheral impact parameter dependence in the region $s>s_R$.
Such dependence is a manifestation of the self--damping of the inelastic channels
at small impact parameters.
The function $h_{inel}(s,b)$ reaches its maximum
 value at $b=R(s)$,
 while the elastic scattering (due to reflection) occurs effectively at smaller values
of impact parameter, i.e.
$\langle b^2 \rangle_{el}<\langle b^2 \rangle_{inel}$. Note that
\[
\langle b^2 \rangle_{i}= \frac{1}{\sigma_i}\int b^2
d\sigma_i \equiv \frac{1}{\sigma_i}\int_0^\infty b^2
\frac{d\sigma_i}{db^2}db^2,
\]
 where $i=el,inel$.

At the  values of energy $s>s_R$ the equation $U(s,b)=1$ has a solution in the
physical region of impact parameter values, i.e. $S(s,b)=0$ at $b=R(s)$. This line is shown
in  the $s$ and $b$ plane in Fig. 1 alongside with the
regions where elastic $S$-matrix has positive and negative values.
   \begin{figure}[hbt]
\begin{center}
\includegraphics[scale=0.5]{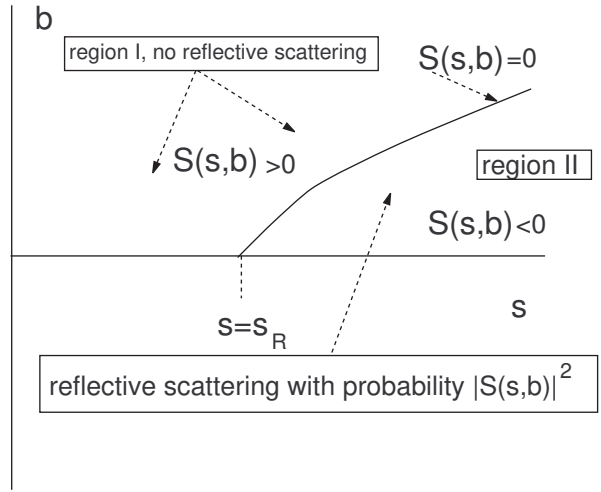}
\caption{\small\it Regions of positive (absorptive scattering) and negative values
(absorptive and reflective scattering) of the
function $S(s,b)$ in the $s$ and $b$ plane.}
\end{center}
\end{figure}
The dependence of $S(s,b)$ on impact
 parameter $b$ at fixed energies (in the region $s>s_R$) is depicted on
 Fig. 2.
\begin{figure}[hbt]
\begin{center}
\includegraphics[scale=0.4]{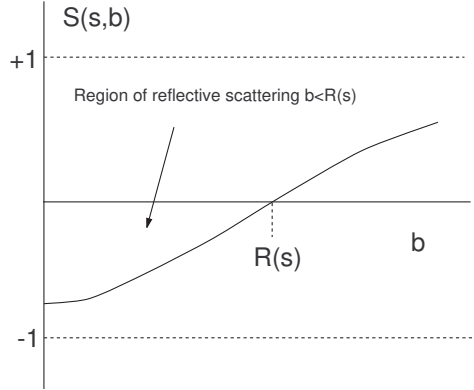}
\caption{\small\it Qualitative impact parameter dependence  of the
function $S(s,b)$ for  the energies  $s>s_R$.}
\end{center}
\end{figure}

The probability of reflective scattering at $b<R(s)$ and $s> s_R$ is determined by the magnitude
 of $|S(s,b)|^2$; this probability is equal to zero at $s\leq s_R$ and $b\geq R(s)$ (region I on Fig.1).
 The form of $U(s,b)$
 depends on the particular model assumptions,
but for our qualitative
 purposes it
is sufficient that it increases with energy in a power-like way
 and decreases with impact parameter
like a linear exponent.
 The dependence of $R(s)$
 is determined then by the logarithmic functional dependence $R(s) \sim \frac{1}{M}\ln s$ , this dependence
 is consistent with analytical properties of the resulting elastics scattering amplitude in
  the complex $t$-plane and mass $M$ can be related to the pion mass.

 Thus, at the energies $s> s_R$  reflective scattering will mimic presence of repulsive core in
 hadron and meson interactions as well.
  Presence of the reflective scattering can be
 accounted for  using van der Waals method (cf. \cite{cleym}).
 This approach was used originally for description of the  fluids behavior starting from
 the gas approximation by means of
  taking into account the nonzero size of molecules.
\section{Effects of the reflective scattering mode for nuclear collisions}
Consider central collision of two identical nuclei having $N$ hadrons in total with center of mass energy
$\sqrt{s}$ per nucleon and calculate
hadron density $n_R(T,\mu)=N/V$ in the initial state at given
 temperature $T$ and baryochemical potential $\mu$ in the presence of
reflective scattering.
The effect of the reflective scattering of
  hadrons is equivalent  to decrease of the volume of the available  space
  which the hadrons are able to occupy in the case when reflective scattering is absent.
  Thus followings to van der Waals method, we must then replace volume $V$ by $V-p_R(s)V_R(s)\frac{N}{2}$,
  i.e. we should write
  \[
  n(T,\mu)=\frac{N}{V-p_R(s)V_R(s)\frac{N}{2}},
\]
where $n(T,\mu)$ is hadron density  without account for reflective scattering and
 $p_R(s)$ is the averaged over volume $V_R(s)$ probability of reflective scattering:
\[
p_R(s)=\frac{1}{V_R(s)}\int_{V_R(s)}|S(s,r)|^2 d^3x.
\]
The volume $V_R(s)$ is determined by the radius of the reflective scattering.
Here we  assume spherical symmetry of hadron interactions,
i.e. we replace impact parameter $b$ by $r$ and approximate the volume $V_R(s)$ by
$V_R(s)\simeq (4\pi/3)R^3(s)$. Hence, the density $n_R(T,\mu)$  is connected
with corresponding density in the approach without reflective scattering $n(T,\mu)$
 by the following relation
\[
  n_R(T,\mu)=\frac{n(T,\mu)}{1+\alpha(s)n(T,\mu)},
\]
where $\alpha(s)={p_R(s)V_R(s)}/{2}$. Let us now estimate change of the function
$n_R(T,\mu)$ due to the presence of reflective scattering. We can
approximate $p_R(s)$ by the value of $|S(s,b=0)|^2$ which tends to
unity at $s\to\infty$. It should be noted that the value
$\sqrt{s_R}\simeq 2$ $TeV$ \cite{sr}. Below this energy there is
no reflective scattering, $\alpha(s)=0$ at $s\leq s_R$, and
therefore corrections to the hadron density are absent. Those
corrections are small when the energy is not too much higher than
$s_R$. At $s\geq s_R$ the value of $\alpha(s)$ is positive,
and presence of reflective
scattering diminishes hadron density. We should expect that this
effect would already be  noticeable at the LHC energy $\sqrt{s}\simeq 5$
TeV in $Pb+Pb$ collisions. At very high energies ($s\to\infty$)
\[
n_R(T,\mu)\sim 1/\alpha(s)\sim M^3/\ln ^3 s.
\]
This  limiting dependence for the hadron density  appears due to
 the presence of the reflective scattering which results in similarity of head-on hadron collisions
with  scattering of hard spheres. It
can be associated with  saturation of the Froissart-Martin bound for the total cross-section. It should be noted
that this dependence has been obtained under assumption on spherical symmetry of hadron interaction region.
Without this assumption, limiting dependence of the hadron density in transverse plane can only be obtained,
i.e. transverse plane density of hadrons would have then the following behavior
\[
n_R(T,\mu)\sim M^2/\ln ^2 s.
\]
\section*{Conclusion}
To conclude, we would like to note that the lower densities of hadron matter are needed
for percolation (and transition to the deconfined state) in the presence of reflective scattering.
It might be useful to note that the rescattering processes
  are also affected by the reflective scattering.
Reflective scattering would lead to noticeable effects at the LHC energies and beyond and
could help in searches of the
deconfined state and studies of properties of transition mechanism to this state of matter which might
proceed by means of percolation. Thus, it will affect description of
initial state dynamics in nuclear
interactions at the LHC energies
 by introducing notion of limiting density of strongly interacting matter at respective energies.
 The appearance  of limiting density dependent on energy takes place only
  at very high energies and has a dynamical origin related to unitarity saturation.
\section*{Acknowledgements}
We are grateful to V.A. Petrov for the interesting discussions of reflective scattering mode.

\end{document}